\documentstyle[epsfig,12pt]{elsart}
\input epsf
\newcommand{\be}[1]{\begin{equation} \label{(#1)}}
\newcommand{\ee}{\end{equation}}
\newcommand{\ba}[1]{\begin{eqnarray} \label{(#1)}}
\newcommand{\ea}{\end{eqnarray}}
\newcommand{\nn}{\nonumber}
\newcommand{\rf}[1]{(\ref{(#1)})}
\newcommand{\lsim}{\;\raisebox{-0.9ex}{$\textstyle\stackrel{\textstyle<}
           {\sim}$}\;}
\def\fig#1{{Fig. (\ref{#1})}}
\def \lsim {\mbox{${}^< \hspace*{-7pt} _\sim$}}

\def\rp{$R_p \hspace{-1em}/\;\:$ }

\def \znbb {\beta\beta_{0\nu}}

\def \emass {\langle m_{\nu} \rangle}
\begin{document}

\begin{flushright}hep-ph/0002264\\ FTUV/00-14\\ IFIC/00-15 \end{flushright}  

\begin{frontmatter}
  
  \title{Bilinear R-parity violating SUSY: Neutrinoless double beta
    decay in the light of solar and atmospheric neutrino data}

\author{M. Hirsch\thanksref{mahirsch}}
\author{J. C. Rom\~ao\thanksref{romao}}
\author{J.~W.~F. Valle\thanksref{valle}}
\address{(a) Instituto de F\'{\i}sica Corpuscular 
-- C.S.I.C. \\ Departamento de F\'{\i}sica Te\`orica, 
Universitat of Val\`encia, \\ Edificio Institutos de Paterna,
Apartado de Correos  2085 \\ 46071 Val\`encia}
\address{(b) Instituto Superior T\'ecnico,
  Departamento de F\'{\i}sica\\
  A. Rovisco Pais 1, 1049-001 Lisboa, Portugal}

\thanks[mahirsch]{(a) mahirsch@neutrinos.uv.es}
\thanks[romao]{ (b) romao@gtae3.ist.utl.pt}
\thanks[valle]{ (a) valle@neutrinos.uv.es}

\begin{abstract}
  
  Neutrinoless double beta ($\znbb$) decay is considered within
  bilinear R-parity breaking supersymmetry, including the full
  one-loop corrections to the neutrino-neutralino mass matrix.
  Expected rates for $\znbb$ decay in this model are discussed in
  light of recent atmospheric and solar neutrino data. We conclude
  that (a) tree-level calculations for $\znbb$ decay within the
  bilinear model are not reliable in the range of parameters preferred
  by current solar and atmospheric neutrino problems. And (b) if the
  solar and atmospheric neutrino problems are to be solved within
  bilinear R-parity violating SUSY the expected rates for $\znbb$
  decay are very low; the effective Majorana neutrino mass at most
  $0.01$ eV and typical values being one order of magnitude lower.
  Observing $\znbb$ decay in the next round of experiments therefore
  would rule out the bilinear R-parity violating supersymmetric model
  as an explanation for solar and atmospheric neutrino oscillations,
  as well as any hierarchical scheme for neutrino masses, unless new
  neutrino interactions are present.
\end{abstract}
\end{frontmatter}

\section{Introduction}

Neutrino physics has entered a new era recently with the announcement
by the Super-Kamiokande collaboration of rather conclusive evidence for
neutrino oscillations \cite{Fukuda:1998mi} in atmospheric neutrino
measurements. This experiment, together with the oscillation
interpretation of the long-standing solar neutrino puzzle
\cite{Smy:1999tt} now provides important information on neutrino
masses and mixings and may-be the first look to physics beyond the
standard model~\cite{MSW99,atm99}.

However, neutrino oscillation experiments, while being extremely
valuable, can not answer two fundamental questions in neutrino
physics. First, they are only sensitive to mass squared differences
and thus can not fix the overall mass scale of neutrinos. And, second,
due to the V-A nature of the weak interaction neutrino oscillations
can not distinguish in practice between Dirac and Majorana neutrinos.
\footnote{The oscillations which are Dirac--Majorana--sensitive must
 violate lepton number by two units and are helicity suppressed
  \protect\cite{Schechter:1981gk}}
Other experiments on neutrino masses are needed in order to
reconstruct the neutrino mass matrix. Neutrinoless double beta decay
is a prominent example of such kind of experiments.

Neutrinoless double beta ($\znbb$) decay has for a long time been 
known as a sensitive probe for physics beyond the standard model 
(SM). Non-observation of $\znbb$ decay has been used to derive 
stringent limits on various extensions of the SM, like, for example, 
left-right symmetric models \cite{LR}, leptoquarks \cite{LQ} and 
supersymmetry \cite{TrRPV,FKS98,HV99}. However, $\znbb$ decay has 
yet to be observed experimentally. 

Although there might exist a variety of mechanisms inducing $\znbb$ decay 
in gauge theories, one can show that whatever the leading mechanism 
is at least one of the neutrinos will be a Majorana particle \cite{SV82}. 
The observable in $\znbb$ decay, the effective Majorana neutrino mass, 
is in general a superposition of different mass eigenstates:

\be{defmass}
\emass = \sum_j' U_{ej}^2 m_j,
\ee
where $U_{ej}$ characterizes the couplings of the mass-eigenstate
neutrinos to the electron in the charged current and the prime
indicates that the sum runs over light mass eigenstates only. If
neutrinos have non-zero mass, also non-zero mixing among them has to
be expected, so that in general $\emass$ does not coincide with the
electron neutrino mass probed in tritium beta decay.

Currently the most stringent experimental bound \cite{hdmo99} gives 
an upper limit of the order of $\emass \le {\cal O}(0.2-0.5)$ $eV$. 
There exist two independent proposals for future experiments which 
might improve the sensitivity on $\emass$ by up to one order of 
magnitude or more \cite{prop,prop2}.

Here, we concentrate on the calculation of expected rates for $\znbb$
decay within bilinear R-parity violating (BRPV) SUSY. While $\znbb$
decay has already been considered in the literature before within the
explicit BRPV SUSY model \cite{FKS98,BFK98,HV99}, it has so--far only been
treated in lowest order of perturbation theory considering the
neutrino-neutralino mass matrix only at the tree-level approximation.
Here, we take into account the full one-loop corrections to the
neutrino-neutralino mass matrix and especially concentrate on those
regions in parameter space in which the model can solve simultaneously
the solar and atmospheric neutrino problems \cite{RPlett}.

We have found that there exist important regions in the parameter
space of the model -- namely those where the BRPV SUSY model can
account for the solar neutrino anomaly through matter--enhanced
oscillations -- where the tree-level estimates for $\znbb$ decay fail
rather badly.  Thus the one-loop corrections considered here play an
important role in BRPV SUSY. Their inclusion is definitely necessary in 
order to predict reliably the effective Majorana neutrino mass relevant 
for $\znbb$ decay in a way consistent with the results from present 
oscillation experiments.

This paper is organized as follows. In the next section we set up the
notations and discuss the model at tree-level. Then, we outline
briefly the extension of the calculation including the one-loop
corrections. Further details for these can be found in \cite{RPlong}.
Section 4 discusses our numerical results.

\section{Bilinear R-parity violation and neutrino mass at tree-level}

In the following we use conventions such that in the limit were the
R-parity violating parameters vanish the usual MSSM notations of refs.
\cite{HabKaneGun} are recovered. For the BRPV case see ref. 
\cite{Diaz:1998xc,chargedhiggs} for the conventions we adopt. The
supersymmetric Lagrangian is specified by the superpotential $W$ given
by
\be{SuperPot}
W=\varepsilon_{ab}\left[ 
 h_U^{ij}\widehat Q_i^a\widehat U_j\widehat H_u^b 
+h_D^{ij}\widehat Q_i^b\widehat D_j\widehat H_d^a 
+h_E^{ij}\widehat L_i^b\widehat R_j\widehat H_d^a 
-\mu\widehat H_d^a\widehat H_u^b 
+\epsilon_i\widehat L_i^a\widehat H_u^b\right] 
\ee
where $i,j=1,2,3$ are generation indices, $a,b=1,2$ are $SU(2)$
indices, and $\varepsilon$ is a completely antisymmetric $2\times2$
matrix, with $\varepsilon_{12}=1$. The symbol ``hat'' over each letter
indicates a superfield, with $\widehat Q_i$, $\widehat L_i$, $\widehat
H_d$, and $\widehat H_u$ being $SU(2)$ doublets with hypercharges
$\frac{1}{3}$, $-1$, $-1$, and $1$ respectively, and $\widehat U$,
$\widehat D$, and $\widehat R$ being $SU(2)$ singlets with
hypercharges $-{\textstyle{4\over 3}}$, ${\textstyle{2\over 3}}$, and
$2$ respectively. The couplings $h_U$, $h_D$ and $h_E$ are $3\times 3$
Yukawa matrices, and $\mu$ and $\epsilon_i$ are parameters with units
of mass. The last term in eq. \rf{SuperPot} is the only $R$--parity
violating term.

Supersymmetry breaking is parameterized with a set of soft 
supersymmetry breaking terms,
\ba{Vsoft}
V_{soft}&=& 
M_Q^{ij2}\widetilde Q^{a*}_i\widetilde Q^a_j+M_U^{ij2} 
\widetilde U_i\widetilde U^*_j+M_D^{ij2}\widetilde D_i 
\widetilde D^*_j+M_L^{ij2}\widetilde L^{a*}_i\widetilde L^a_j+ 
M_R^{ij2}\widetilde R_i\widetilde R^*_j \nonumber\\ 
&&\!\!\!\!+m_{H_d}^2 H^{a*}_d H^a_d+m_{H_u}^2 H^{a*}_u H^a_u- 
\left[\half M_s\lambda_s\lambda_s+\half M\lambda\lambda 
+\half M'\lambda'\lambda'+h.c.\right] \nn \\ 
&&\!\!\!\!\!\!\!\!\!\!\!\!\!\!\!\!\!\!\!\!+\varepsilon_{ab}\left[ 
A_U^{ij}\widetilde Q_i^a\widetilde U_j H_u^b 
+A_D^{ij}\widetilde Q_i^b\widetilde D_j H_d^a 
+A_E^{ij}\widetilde L_i^b\widetilde R_j H_d^a 
-B\mu H_d^a H_u^b+B_i\epsilon_i\widetilde L_i^a H_u^b\right] 
\ea

and again, the last term in eq. \rf{Vsoft} is the only R--parity
violating term. The bilinear term in \rf{Vsoft} leads in the neutral
part of the scalar potential to terms linear in the sneutrino fields.
Thus, in general the sneutrino fields acquire VeVs. This in turn leads
to mixing between the gaugino and lepton as well as to mixing between
the scalar leptons and the Higgs fields
\cite{chargedhiggs,deCampos:1995av}.

For our purposes the most important aspect is the neutrino-neutralino 
mixing, since it leads at tree-level to one massive neutrino state. 
In the basis, ${\Psi'_{0}}^T =$ $(\psi_{L_1}^1, \psi_{L_2}^1, \psi_{L_3}^1,$ 
$-i\lambda', -i \lambda_3,\psi_{H_1}^1,\psi_{H_2}^2)$ the 
neutrino-neutralino mass matrix at tree-level can be written 
as:

\be{nmm}
{\cal M}_0 =  \left(
                    \begin{array}{cc}
                    0 & m \\
                    m^T & {\cal M}_{\chi^0} \\
                    \end{array}
              \right).
\ee
Here, the sub-matrix $m$ contains entries from the bilinear \rp 
parameters,

\be{bnmm}
m =   \left(
            \begin{array}{cccc}
     -\frac{1}{2}g' v_e & \frac{1}{2}g v_e & 0 & \epsilon_e \\
-\frac{1}{2}g' v_{\mu} & \frac{1}{2}g v_{\mu} & 0 & \epsilon_{\mu} \\
        -\frac{1}{2}g' v_{\tau} & \frac{1}{2}g v_{\tau} & 
          0 & \epsilon_{\tau} \\
                    \end{array}
              \right),
\ee
where $v_i := \langle {\tilde \nu}_{i} \rangle$ and ${\cal
  M}_{\chi^0}$ is the MSSM neutralino mass matrix, given by,

\be{MSSMnm}  
{\cal M}_{\chi^0} =  \left(
                        \begin{array}{cccc}
 M_1 & 0   & -\frac{1}{2}g' v_d &  \frac{1}{2} g' v_u  \\
 0   & M_2 &  \frac{1}{2}g  v_d & -\frac{1}{2} g  v_u  \\
  -\frac{1}{2} g' v_d &  \frac{1}{2} g v_d & 0 & -\mu  \\
   \frac{1}{2} g' v_u & -\frac{1}{2} g v_u & -\mu & 0 \\
 \end{array}
                     \right).
\ee

There are two interesting aspects concerning ${\cal M}_0$. First,
${\cal M}_0$ has such a texture that at tree-level only one neutrino
gets a non-zero mass \cite{onlyone}, leaving two massless (but mixed)
states in the spectrum. And second, at tree-level the neutrino mass is
strictly proportional to the ``alignment vector'' $|{\vec \Lambda}|^2$,
where,

\be{deflam}
{\vec \Lambda} := {\vec \epsilon} v_d + {\vec v} \mu.
\ee

Thus, at tree-level the individual $\epsilon_i$ and $v_i$ are not
constrained neither by the neutrino mass measurements nor by
neutrinoless double beta decay, as long as they are sufficiently
aligned. However, we would like to stress (more details below) that
this is a pure tree-level result. Once the calculation is improved to
one-loop order current experimental hints on solar and atmospheric
neutrino oscillations provide rather stringent constraints not only on
${\vec \Lambda}$, but also on the individual BRPV parameters,
$\epsilon_i$ and $v_i$.

Assuming that $m \ll {\cal M}_0$ one can find \cite{FKS98,HV99} a
simple formula relating the effective Majorana neutrino mass to the
supersymmetric parameters:

\be{emassBRPV}
\emass \simeq \frac{2}{3} \frac{g^2 M_2}{det({\cal M}_0)} 
              \Lambda_e^2.
\ee

It has been shown in \cite{HV99} that within BRPV the contribution
from $\emass$ as given above is the dominant source for $\znbb$ decay.
In the following we will concentrate on this BRPV mass mechanism only,
improving it by taking into account the one-loop corrections to the
neutrino-neutralino mass matrix.

\section{One-loop corrections to the neutrino-neutralino mass matrix}

%
%
As we have seen the effective neutrino mass matrix has a projective
structure, such that only one neutrino gets a mass at tree-level. As a
result for a realistic description of the neutrino spectrum one has to
improve the calculation to 1-loop order. \footnote{With two massless
  neutrinos, one angle of the neutrino sector of the theory could be
  rotated away.  Thus a discussion of the predictions of the theory
  for the solar angle is meaningless at tree-level.} A shortened
description is given below, for a complete listing of all necessary
couplings etc.  see ref.~\cite{RPlong}.  However, most important for
the understanding of the importance of the loops is the fact that
these contributions explicity break the projectivity of the tree-level
mass matrix, incorporating contributions which are proportional to the
$\epsilon_i$ themselves, as we will show explicitly below. In
contrast, as discussed above, the tree-level mass matrix is sensitive
only to ${\vec \Lambda}$.

The full neutrino-neutralino mass matrix including the 1-loop
corrections is given by

\be{full}\nn
M_{ij}  = M^{tree}_{ij} + \Delta M_{ij},
\ee
where $\Delta M_{ij}$ are the 1-loop corrections defined by 
\be{1loopMass}\nn
\Delta M_{ij}  = 
               \frac{1}{2} \Big(
                \Pi_{ij}(p_i^2) + \Pi_{ij}(p_j^2) 
                -  m_{\chi^0_i} \Sigma_{ij}(p_i^2) - 
                  m_{\chi^0_j} \Sigma_{ij}(p_j^2) \Big)
\ee
%
%
%
where $\Sigma_{ij}$ and $\Pi_{ij}$ are self-energies. 
There are three simple topologies of relevant Feynman diagrams
contributing to the neutrino-neutralino mass matrix \cite{RPlong}.
\footnote{For a complete description see ref.~\protect\cite{RPlong}}.
Here, ${\overline {DR}}$ signifies the minimal dimensional reduction
subtraction scheme and $\mu_R$ is the renormalization scale. As
pointed out in \cite{RPlong} the inclusion of the tadpole diagram is
essential in order to obtain gauge invariance of the calculation.

\begin{figure}[h]
\setlength{\unitlength}{1mm}
\begin{picture}(50,40)
\put(-40,-210)
{\mbox{\epsfig{figure=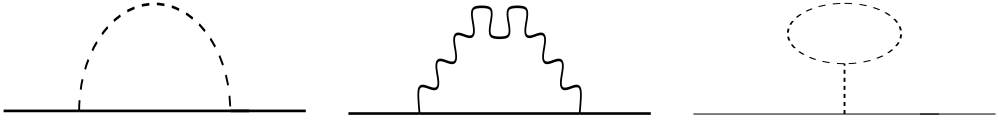,height=28.0cm,width=22.0cm}}}
\end{picture}
\caption[]{ Simple one-loop topologies contributing to the
  neutralino-neutrino mass matrix, see text.}
\label{topologies}
\end{figure}

\bigskip Figure 1 shows the relevant Feynman graphs. Internal
particles in the scalar self-energies can be either ($q-{\tilde q}$),
(charged scalars-charginos) or (neutral scalars-neutralinos), for the
gauge loops it can either be ($W^{\pm}-$charginos) or
($Z^0-$neutralinos). Which of the loops is most important depends both
on parameters and whether one considers the heavy states
(``neutralinos'') or the light states (``neutrinos'').  Here we
concentrate on the ``neutrino'' states. For these only the ($d-{\tilde
  d}$), (charged scalars-charginos) and ($W^{\pm}-$charginos)
combinations do indeed contribute.  For large values of $\tan\beta$
generally the ($d-{\tilde d}$) loops are most important. We will
therefore concentrate on this loop in the following, noting in passing
that the basic structure of all the self-energies are the same and can
be found by replacing internal masses and couplings correspondingly
\cite{RPlong}.

It is interesting to note that the tree-level result of neutrino
masses being strictly proportional to $|{\vec \Lambda}|^2$ is no
longer valid once the one-loop contributions are taken into account.
This can be shown for example for the down-type squark loops, for
which $\Pi_{ij}(p_i^2)$ and $\Sigma_{ij}(p_i^2)$ are given by,

\be{defpi}
\Pi_{ij}(p_i^2) = \frac{-1}{16\pi^2} \sum_{k,s} 
                  \Big( {\cal O}^{nds}_{L,jks} {\cal O}^{dns}_{L,kis} 
                      + {\cal O}^{nds}_{R,jks} {\cal O}^{dns}_{R,kis} \Big) 
                        m_k B_0(m_i^2,m_k^2,m_s^2) 
\ee

\be{defsig}
\Sigma_{ij}(p_i^2) = \frac{-1}{16\pi^2} \sum_{k,s} 
                  \Big( {\cal O}^{nds}_{R,jks} {\cal O}^{dns}_{L,kis} 
                      + {\cal O}^{nds}_{L,jks} {\cal O}^{dns}_{R,kis} \Big) 
                         B_1(m_i^2,m_k^2,m_s^2) 
\ee
where $B_0$ and $B_1$ are Passarino-Veltman functions
\cite{Passarino:1979jh}, $m_k$ and $m_s$ are the down-type quark,
down-type squark masses and the various ${\cal O}$ are
neutralino-quark-squark couplings, in our notation given by,

\be{defOL}
{\cal O}^{\rm dns}_{L ijk} = -\frac{2}{3}(\frac{g}{\sqrt{2}})
        \tan\theta_W {\cal N}_{j5}^* {\bf R}^{\tilde d^*}_{k,m+3}
        {\bf R}^{d}_{{\bf R}i,m}
        - (h_d)_{ml}{\bf R}^{\tilde d^*}_{k,m}
          {\bf R}^{d}_{{\bf R}i,l}{\cal N}_{j7}^*
\ee
\be{defOR}
{\cal O}^{\rm dns}_{R ijk} = (\frac{g}{\sqrt{2}})( {\cal N}_{j6}
       - \frac{1}{3}\tan\theta_W {\cal N}_{j5} )
        {\bf R}^{\tilde d^*}_{k,m}
        {\bf R^*}^{d}_{{\bf L}m,i}
        - (h_d^*)_{ml}{\bf R}^{\tilde d^*}_{k,l+3}
          {\bf R^*}^{d}_{{\bf L}m,i}{\cal N}_{j7}
\ee
where the $h_d$ denote the down-type Yukawa couplings and ${\cal
  O}^{\rm nds}_{L ijk}= \left( {\cal O}^{\rm dns}_{R jik} \right)^*$
and ${\cal O}^{\rm nds}_{R ijk}= \left({\cal O}^{\rm dns}_{L jik}
\right)^*$.  The rotation matrices ${\bf R}^{d}$ and ${\bf R}^{\tilde
  d}$ are the ones which diagonalize the quark and squark mass
matrices, respectively, while ${\cal N}$ diagonalizes the
neutralinos/neutrinos.

That terms proportional to $\epsilon_i$ survive in eq. \rf{defpi} is
most easily seen assuming the BRPV parameters are small, as suggested
by the present indications from solar and atmospheric neutrino data. 
Then one can block-diagonalize the neutrino-neutralino mass matrix
perturbatively at tree level in terms of the expansion parameter $\xi
= m \cdot {\cal M}_{\chi^0}^{-1}$ \cite{Schechter:1982cv} as,

\ba{xiapprox}
{\cal N}^* & = & \left(\begin{array}{cc}
V_\nu^T(1 -{1 \over 2}\xi \xi^{\dagger} ) & -V_\nu^T\xi \\
N^* \xi^{\dagger} & N^* ( 1 -{1 \over 2}\xi^\dagger \xi)
\end{array}\right)
\ea
where $N^*$ is the matrix diagonalizing the MSSM part of the
neutralino mass matrix and $V_\nu^T$ describes the mixing of neutrinos
among themselves.

The full form for the expansion matrix $\xi$ can be found, for
example, in \cite{HV99}. For our purposes it suffices to state that in
the limit ${\vec \Lambda} \equiv 0$ the matrix $V_\nu^T$ is diagonal,
and all elements of $\xi$ vanish except $\xi_{i3}$, which take the
simple form,

\be{xi3}
\xi_{i3} = - \frac{\epsilon_i}{\mu} 
\ee

Inserting this result for ${\vec \Lambda} \equiv 0$ and for simplicity
considering only $i,j =1,2,3$, ($\Sigma_{ij}$ vanishes for $i,j
=1,2,3$ in this limit) $\Pi_{ij}$ can be written as,

\be{simpi}
\Pi_{ij}(p_i^2) = \frac{-1}{16\pi^2} \frac{\epsilon_i \epsilon_j}{\mu^2}
                  \sum_{k,s} 
                  \Big( 
                  {\bf R}^{\tilde d}_{s,k+3}{\bf R}^{\tilde d^*}_{s,k}
                   + h.c.
                   \Big) 
                   |(h_d)_{kk}|^2 m_k B_0(m_i^2,m_k^2,m_s^2), 
\ee
where, for simplicity, we have assumed that $h_d$ is diagonal.  Eq.
\rf{simpi} demonstrates that the entries in $\Pi_{ij}$ in the
``neutrino sector'' are proportional to $\epsilon_i \epsilon_j$.  This
shows explicitly that in the limit where the tree-level neutrino mass
vanishes the loop contributions do not and can, potentially, be rather
important. 
%
%
Moreover, from this example we can draw two conclusions. First, 1-loop 
contributions break the projectivity of the mass matrix 
($m^{tree}_{ij} \sim \Lambda_i\Lambda_j$ at tree-level) and thus 
the degeneracy of the two lightest states is lifted. And, second, 
the size of the ratio of the 1-loop to the tree-level entries of 
the mass matrix should be controled mainly by the quantity 
$|{\vec \epsilon}|^2/|{\vec \Lambda}|$. \footnote{In the numerical 
calculation we have found that this is indeed the case. However, 
the loops depend also strongly on $\tan\beta$, because large $\tan\beta$ 
leads to large Yukawa couplings in the down sector, and as shown in 
eq. \rf{simpi} the 1-loop entries strongly depend on $h_d$. Numerically, 
variations of other SUSY parameters have been found to be much less 
important.}

%
%

\section{Numerical results}

In our numerical study we assume unification at a scale $Q = M_U$ with
standard minimal supergravity boundary conditions,
\ba{gut}
&&A_t=A_b=A_{\tau}\equiv A\:,\nonumber\\
&&B=B_i=A-1 \:,\nonumber\\
&&m_{H_d}^2=m_{H_u}^2=M_{L_i}^2=M_{R_i}^2=m_0^2 \\
&&M_{Q_i}^2=M_{U_i}^2=M_{D_i}^2=m_0^2\:,\nonumber\\
&&M_3=M_2=M_1=M_{1/2}\,.\nonumber
\ea

We run the RGE's from the unification scale $M_U \sim 2\times10^{16}$
GeV down to the weak scale, giving random values to the fundamental
parameters at the unification scale.  We then check that the numerical
values obtained from the RGE running correctly break electroweak
symmetry.  Moreover, we accept only those points for further study,
which fulfill phenomenological constraints from negative Higgs and
SUSY particle searches at accelerators \cite{Caso:1998tx}.

Although this procedure is not essential for the calculation 
of the neutrino masses in the model, it allows us to reduce 
the number of free parameters considerably and can be viewed 
as a test for self-consistency of the parameter ranges under 
consideration.

For the \rp parameters, we use the constraints from solar and
atmospheric neutrinos found in \cite{RPlett,RPlong}. These two sets of
measurements imply that BRPV parameters have to be small, i.e.
$|\epsilon|$ and $|\Lambda|$ should be smaller than ${\cal O}$(GeV)
and ${\cal O}(0.2 GeV^2)$ respectively for typical MSSM parameters
smaller than, say $1$ TeV.  \footnote{Although smaller than usual
  supersymmetric parameters, such a suppression might be actually
  expected in scenarios with radiative R-parity breaking
  ref.~\protect\cite{RPlong,Diaz:1998xc}} Moreover, measurements of
(or limits on) neutrino angles fix (or yield limits) on ratios of
R-parity breaking parameters.
\begin{figure}
\setlength{\unitlength}{1mm}
\begin{picture}(50,70)
\put(-30,-90)
{\mbox{\epsfig{figure=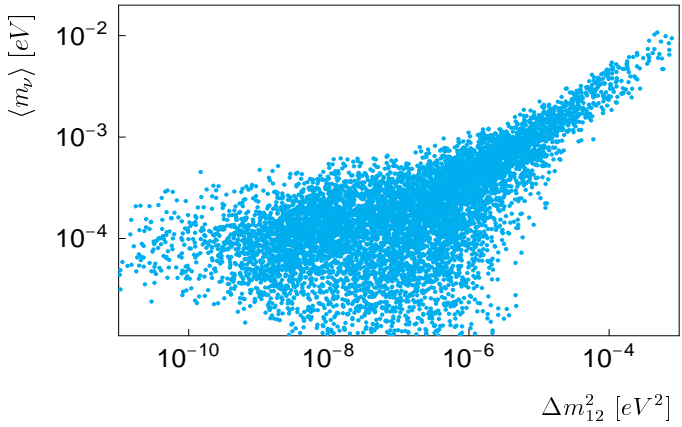,height=22.0cm,width=18.0cm}}}
\end{picture}
\caption[]{Effective Majorana neutrino mass as a function of 
$\Delta m^2_{12}$ for data points which have 
$\sin^2(2\theta_{sol}) \ge 0.6$ and solve the atmospheric neutrino 
problem.}
\label{dbdvdelm12}
\end{figure}
\begin{figure}
\setlength{\unitlength}{1mm}
\begin{picture}(50,70)
\put(-30,-90)
{\mbox{\epsfig{figure=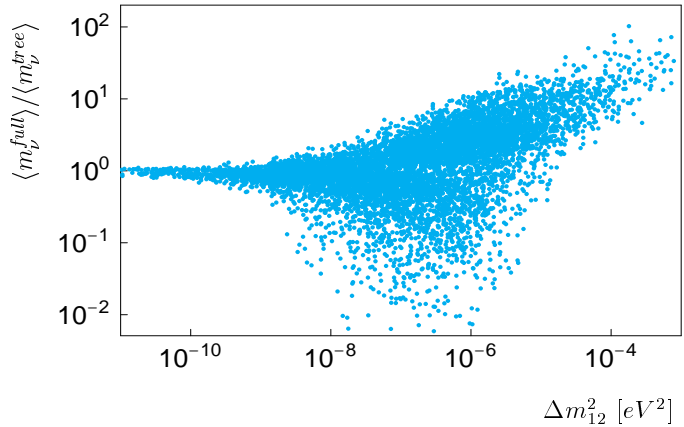,height=22.0cm,width=18.0cm}}}
\end{picture}
\caption[]{Ratio of 1-loop corrected effective Majorana 
  neutrino mass to its tree-level value as a function of $\Delta
  m^2_{12}$ for data points which have $\sin^2(2\theta_{sol}) \ge 0.6$
  and solve the atmospheric neutrino problem.}
\label{dbdratvdelm12}
\end{figure}
Here we summarize these restrictions as follows \cite{RPlong}. The
atmospheric neutrino measurements require $\Lambda_{\mu} \simeq
\Lambda_{\tau}$, whereas the negative results from the CHOOZ
\cite{chooz} and Palo Verde reactor \cite{PaloVerde} experiments
require that $\Lambda_{e}$ should be smaller than $\Lambda_{e} \le
{\cal O}(0.3) \sqrt{\Lambda_{\mu}^2+\Lambda_{\tau}^2}$.  The solar
neutrino problem can be either solved with relatively large mixing
(LMA-MSW or vacuum oscillations), which implies that all $\epsilon_i$
should be similar, or by small mixing (the SMA-MSW solution), the
latter implying $\epsilon_e \sim {\rm (few)} 10^{-2}
\epsilon_{\mu,\tau}$.

We have determined the expected values of $\emass$ as a function of
$\Delta m^2_{12}$ for about $10^4$ calculated points, which solve the
atmospheric neutrino problem. Predicted values of $\emass$ are rather
small, reaching at most $10^{-2}$ [eV] for the large mixing solution
(LA-MSW) of the solar neutrino problem, as can be seen from
\fig{dbdvdelm12}. For the case of vacuum oscillations $\emass$ will be
even much smaller, around $10^{-4}$ [eV], as seen from the figure.

Let us now discuss the crucial importance of the loop corrections to
the neutrino masses in this context. In order to do this we have
calculated ratios of $\emass$ including the 1-loop corrections divided
by its tree-level value.  In figure \fig{dbdratvdelm12} we show our
results.  As can be seen, if $\Delta m^2_{12}$ lies in the range
required for vacuum (or just-so) oscillations the tree-level and the
1-loop improved $\emass$ are rather similar, whereas for larger
$\Delta m^2_{12}$ in the MSW range one has a substantial change from
the tree-level result.  Thus, tree-level calculations of $\emass$ are
certainly not accurate in this case, and the 1-loop corrections
considered here play an essential role.

\begin{figure}
\setlength{\unitlength}{1mm}
\begin{picture}(50,70)
\put(-30,-90)
{\mbox{\epsfig{figure=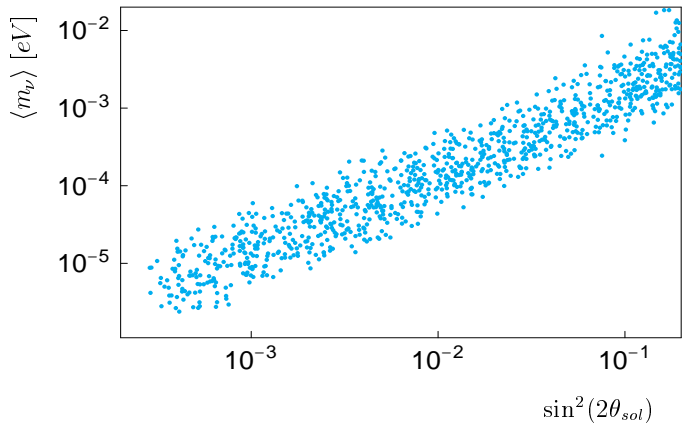,height=22.0cm,width=18.0cm}}}
\end{picture}
\caption[]{Expected 1-loop corrected effective Majorana 
  neutrino mass $\emass$  as a function of $\sin^2(2\theta_{sol})$
  for those points which solve the atmospheric neutrino problem.}
\label{dbdvs2thsol}
\end{figure}

Let us now analyze the remaining oscillation possibility to solve the
solar neutrino problem, namely the small-angle MSW solution. In this
case one finds a suppression in the $\znbb$ rate, as can be seen in 
\fig{dbdvs2thsol}. 
This result is easy to understand conceptually, as the $\znbb$ rate
must be given in terms of the only $L_e$ violating parameters in the
model $\Lambda_e$ and $\epsilon_e$, while $\sin^2(2\theta_{sol}) \to
0$ as $\Lambda_e, \epsilon_e \to 0$.

To close this section we mention that, although we have worked within
the framework of a concrete model in which \rp constitutes the origin
for neutrino mass and mixing, our conclusions are more general. In
fact the smallness of effective Majorana neutrino mass $\emass$ holds
in any hierarchical model of neutrino mass, of which our bilinear \rp
breaking model is a particular case. 
%
%
Note, that although it is possible in the BRpV model to have two
neutrinos nearly degenerate once the 1-loop contributions are
included, it is never possible to have all three neutrinos degenerate
\cite{RPlong}.  Moreover, such points are extremely rare in parameter
space and not protected by any symmetry in our model. In {\em
  hierarchical} models, however one expects that the {\em maximum}
allowed value of $\emass$ (which is achieved for the LA-MSW solution)
can be estimated by:

\ba{estmass}
\emass &=& \sum_j' U_{ej}^2 m_j \\ \nn
       &\sim& U_{e2}^2 \sqrt{\Delta m^2_{sol}} + 
              U_{e3}^2 \sqrt{\Delta m^2_{atm}} \hskip3mm
        \lsim \hskip3mm \frac{1}{2} \sqrt{10^{-4} {\rm eV}^2} + 
              0.05  \sqrt{10^{-2} {\rm eV}^2} 
       \hskip3mm \sim\hskip3mm 0.01 {\rm eV},
\ea
which our numerical results confirm for the BRpV model explicitly. 
Note, that eq. \rf{estmass} gives us only an {\em upper bound} on 
$\emass$, but no {\em lower bound} and no prediction for $\emass$.

One interesting way to avoid this upper bound is the possibility of
neutrinos being closely degenerate in mass.  According to our results,
this would be a clear indication that BRpV is {\bf not} the underlying
mechanism for generating the solar and atmospheric neutrino masses.
Another is if other more exotic mechanisms for solving the neutrino
anomalies are entertained, such as flavour changing interactions or
decays \cite{Fornengo:1999zp}.

\section{Summary} 

We have calculated the one-loop corrections to the $\znbb$ decay
observable $\emass$ in bilinear R-parity violating supersymmetry,
following the procedure developed in \cite{RPlong}. Since it has been
shown in \cite{RPlett,RPlong} that the model is able to solve the
solar and atmospheric neutrino problems under certain, relatively
simple assumptions, special emphasis has been put in our analysis on
those ``successful'' regions of parameter space.

There are two main results of this study. First, one-loop corrections
are important for estimating $\znbb$ decay rates in bilinear BRPV SUSY.
This is due to the fact that the model at tree-level has two massless
states in the spectrum. This degeneracy is lifted once the one-loop
corrections are taken into account. Since tree-level and one-loop
masses depend on different combinations of BRPV parameters, which are a
priori unknown, the loop corrections can be easily as big as the tree
level masses. Especially this is true in those parameter ranges, where
the model is able to solve the solar and atmospheric neutrino
problems.

Moreover we show that, if bilinear R-parity violating is indeed the
solution to the solar and atmospheric neutrino problems, than the
expected values of $\emass$ are very small, certainly smaller than
$10^{-2}$ eV, and probably even smaller than $10^{-3}$ eV.

Although this conclusion might appear rather discouraging for the
experimentalists, we would like to stress that, on the other hand,
discovering $\znbb$ decay at a level significantly larger than $\emass
= 10^{-2}$ eV would be sufficient to rule out our model as an
explanation for the atmospheric and solar neutrino problems.  This
conclusion also applies to any \emph{hierarchical} scheme for
neutrino masses. The only possible way this conclusion might be evaded
is to consider the presence of exotic neutrino properties, such as
flavour changing interactions or decays \cite{Fornengo:1999zp}.

\bigskip
\centerline{\bf Acknowledgement}

This work was supported by DGICYT grant PB98-0693 and by the TMR
contract ERBFMRX-CT96-0090. M.H. acknowledges support from 
the European Union's Marie-Curie program under grant No 
ERBFMBICT983000.

\end{document}